\documentclass[10pt,twocolumn,aps,prc,superscriptaddress,preprintnumbers,showpacs,floatfix]{revtex4-1}  
\usepackage{graphicx}
\usepackage{sidecap}
\usepackage{array}
\usepackage{multirow}
\usepackage{subfigure}
\usepackage{color}
\usepackage{lineno}
\usepackage{float}               
\usepackage{placeins}
\usepackage{xspace}        
\usepackage{epsfig}
\usepackage{graphicx}
\usepackage{fancyhdr}   
\usepackage{pslatex}   
\usepackage{amsmath, amsthm, amssymb} 
\usepackage[symbol]{footmisc}
\usepackage[resetlabels]{multibib}
\usepackage{framed}
\usepackage{supertabular}
\usepackage[colorlinks,linkcolor=blue,anchorcolor=red,citecolor=blue]{hyperref}

\newcites{sec}{Reference}

\newcommand{\yr}{$\mathrm{N}_t \times \mathrm{N}_p/\mathrm{N}_d^2$}
\newcommand{\tp}{$\mathrm{N}_t / \mathrm{N}_p$}
\newcommand{\dop}{$\mathrm{N}_d / \mathrm{N}_p$}
\newcommand{\snn}{$\sqrt{s_\mathrm{NN}}$~}
\newcommand{\dndeta}{$dN_{ch}/d\eta$}

\begin{document}

\preprint{}

\title{ Beam Energy Dependence of Triton Production and Yield Ratio ({\yr}) in Au+Au Collisions at RHIC }

\include{author_new}
\affiliation{Abilene Christian University, Abilene, Texas   79699}
\affiliation{Alikhanov Institute for Theoretical and Experimental Physics NRC "Kurchatov Institute", Moscow 117218}
\affiliation{Argonne National Laboratory, Argonne, Illinois 60439}
\affiliation{American University of Cairo, New Cairo 11835, New Cairo, Egypt}
\affiliation{Ball State University, Muncie, Indiana, 47306}
\affiliation{Brookhaven National Laboratory, Upton, New York 11973}
\affiliation{University of Calabria \& INFN-Cosenza, Rende 87036, Italy}
\affiliation{University of California, Berkeley, California 94720}
\affiliation{University of California, Davis, California 95616}
\affiliation{University of California, Los Angeles, California 90095}
\affiliation{University of California, Riverside, California 92521}
\affiliation{Central China Normal University, Wuhan, Hubei 430079 }
\affiliation{University of Illinois at Chicago, Chicago, Illinois 60607}
\affiliation{Creighton University, Omaha, Nebraska 68178}
\affiliation{Czech Technical University in Prague, FNSPE, Prague 115 19, Czech Republic}
\affiliation{National Institute of Technology Durgapur, Durgapur - 713209, India}
\affiliation{ELTE E\"otv\"os Lor\'and University, Budapest, Hungary H-1117}
\affiliation{Frankfurt Institute for Advanced Studies FIAS, Frankfurt 60438, Germany}
\affiliation{Fudan University, Shanghai, 200433 }
\affiliation{University of Heidelberg, Heidelberg 69120, Germany }
\affiliation{University of Houston, Houston, Texas 77204}
\affiliation{Huzhou University, Huzhou, Zhejiang  313000}
\affiliation{Indian Institute of Science Education and Research (IISER), Berhampur 760010 , India}
\affiliation{Indian Institute of Science Education and Research (IISER) Tirupati, Tirupati 517507, India}
\affiliation{Indian Institute Technology, Patna, Bihar 801106, India}
\affiliation{Indiana University, Bloomington, Indiana 47408}
\affiliation{Institute of Modern Physics, Chinese Academy of Sciences, Lanzhou, Gansu 730000 }
\affiliation{University of Jammu, Jammu 180001, India}
\affiliation{Joint Institute for Nuclear Research, Dubna 141 980}
\affiliation{Kent State University, Kent, Ohio 44242}
\affiliation{University of Kentucky, Lexington, Kentucky 40506-0055}
\affiliation{Lawrence Berkeley National Laboratory, Berkeley, California 94720}
\affiliation{Lehigh University, Bethlehem, Pennsylvania 18015}
\affiliation{Max-Planck-Institut f\"ur Physik, Munich 80805, Germany}
\affiliation{Michigan State University, East Lansing, Michigan 48824}
\affiliation{National Research Nuclear University MEPhI, Moscow 115409}
\affiliation{National Institute of Science Education and Research, HBNI, Jatni 752050, India}
\affiliation{National Cheng Kung University, Tainan 70101 }
\affiliation{The Ohio State University, Columbus, Ohio 43210}
\affiliation{Panjab University, Chandigarh 160014, India}
\affiliation{NRC "Kurchatov Institute", Institute of High Energy Physics, Protvino 142281}
\affiliation{Purdue University, West Lafayette, Indiana 47907}
\affiliation{Rice University, Houston, Texas 77251}
\affiliation{Rutgers University, Piscataway, New Jersey 08854}
\affiliation{University of Science and Technology of China, Hefei, Anhui 230026}
\affiliation{South China Normal University, Guangzhou, Guangdong 510631}
\affiliation{Sejong University, Seoul, 05006, South Korea}
\affiliation{Shandong University, Qingdao, Shandong 266237}
\affiliation{Shanghai Institute of Applied Physics, Chinese Academy of Sciences, Shanghai 201800}
\affiliation{Southern Connecticut State University, New Haven, Connecticut 06515}
\affiliation{State University of New York, Stony Brook, New York 11794}
\affiliation{Instituto de Alta Investigaci\'on, Universidad de Tarapac\'a, Arica 1000000, Chile}
\affiliation{Temple University, Philadelphia, Pennsylvania 19122}
\affiliation{Texas A\&M University, College Station, Texas 77843}
\affiliation{University of Texas, Austin, Texas 78712}
\affiliation{Tsinghua University, Beijing 100084}
\affiliation{University of Tsukuba, Tsukuba, Ibaraki 305-8571, Japan}
\affiliation{University of Chinese Academy of Sciences, Beijing, 101408}
\affiliation{Valparaiso University, Valparaiso, Indiana 46383}
\affiliation{Variable Energy Cyclotron Centre, Kolkata 700064, India}
\affiliation{Wayne State University, Detroit, Michigan 48201}
\affiliation{Yale University, New Haven, Connecticut 06520}

\author{M.~I.~Abdulhamid}\affiliation{American University of Cairo, New Cairo 11835, New Cairo, Egypt}
\author{B.~E.~Aboona}\affiliation{Texas A\&M University, College Station, Texas 77843}
\author{J.~Adam}\affiliation{Czech Technical University in Prague, FNSPE, Prague 115 19, Czech Republic}
\author{J.~R.~Adams}\affiliation{The Ohio State University, Columbus, Ohio 43210}
\author{G.~Agakishiev}\affiliation{Joint Institute for Nuclear Research, Dubna 141 980}
\author{I.~Aggarwal}\affiliation{Panjab University, Chandigarh 160014, India}
\author{M.~M.~Aggarwal}\affiliation{Panjab University, Chandigarh 160014, India}
\author{Z.~Ahammed}\affiliation{Variable Energy Cyclotron Centre, Kolkata 700064, India}
\author{A.~Aitbaev}\affiliation{Joint Institute for Nuclear Research, Dubna 141 980}
\author{I.~Alekseev}\affiliation{Alikhanov Institute for Theoretical and Experimental Physics NRC "Kurchatov Institute", Moscow 117218}\affiliation{National Research Nuclear University MEPhI, Moscow 115409}
% ...

\collaboration{STAR Collaboration}\noaffiliation




\begin{abstract}
We report the triton ($t$) production in mid-rapidity ($|y| <$ 0.5) Au+Au collisions at {\snn}= 7.7--200 GeV measured by the STAR experiment from the first phase of the beam energy scan at the Relativistic Heavy Ion Collider (RHIC). The nuclear compound yield ratio ({\yr}), which is predicted to be sensitive to the fluctuation of local neutron density, is observed to decrease monotonically with increasing charged-particle multiplicity ({\dndeta}) and follows a scaling behavior. The {\dndeta} dependence of the yield ratio is compared to calculations from coalescence and thermal models. Enhancements in the yield ratios relative to the coalescence baseline are observed in the 0\%-10\% most central collisions at 19.6 and 27 GeV, with a significance of 2.3$\sigma$ and 3.4$\sigma$, respectively, giving a combined significance of 4.1$\sigma$. The enhancements are not observed in peripheral collisions or model calculations without critical fluctuation, and decreases with a smaller $p_{T}$ acceptance. The physics implications of these results on the QCD phase structure and the production mechanism of light nuclei in heavy-ion collisions are discussed.

\end{abstract}
\pacs{xx.xx.-q}
\maketitle

Quantum Chromodynamics (QCD) is the fundamental theory that describes the strong interaction. One of the main goals of the Beam Energy Scan (BES) program at Relativistic Heavy Ion Collider (RHIC) is to explore the QCD phase structure~\cite{Rajagopal:2000wf,STAR:2010vob}. Lattice QCD calculations indicate that the transition between hadronic matter and the Quark-Gluon Plasma (QGP) is a smooth crossover at vanishing baryon chemical potential ($\mu_{B}$ = 0 MeV)~\cite{Aoki:2006we}, with a transition temperature of about $T_c = 156$ MeV~\cite{HotQCD:2018pds}.
QCD-based models suggest that there could be a first-order phase transition at large baryon chemical potential~\cite{Ejiri:2008xt,Fischer:2018sdj,Fu:2019hdw,Gao:2020fbl}. If theory postulations are correct, the first-order phase transition line would end at a critical point (CP)~\cite{Halasz:1998qr,Stephanov:2004wx,Fukushima:2010bq}. A fundamental question is whether we can experimentally find the CP and pin down its location in the QCD phase diagram~\cite{Gupta:2011wh,STAR:2013gus,Bzdak:2019pkr,Luo:2017faz,STAR:2020tga,STAR:2021iop}. In the BES program, the STAR experiment has measured the energy dependence of observables that are sensitive to the CP and/or first-order phase transition, including pion HBT radii ~\cite{STAR:2014shf,STAR:2020dav}, baryon directed flow~\cite{STAR:2014clz,STAR:2017okv}, net-proton fluctuations~\cite{STAR:2020tga,STAR:2021iop} and intermittency of charged hadrons~\cite{STAR:2023jpm}. Non-monotonic energy dependencies were observed in all of these observables, and the energy ranges where peak or dip structures appear are around {\snn} $\approx$ 7.7--39 GeV. Those intriguing observations are of great interest and more investigation and analysis are required to reach definitive conclusion. 

Light nuclei, such as deuteron ($d$), triton ($t$), helium-3 ($^{3}\mathrm{He}$), are loosely bound objects with binding energies of several MeV. Their production in heavy-ion collisions is an active area of research both experimentally~\cite{Cocconi:1960zz,E814:1994kon,E864:2000auv,Albergo:2002gi,FOPI:2010xrt,ALICE:2015wav,NA49:2016qvu,STAR:2016ydv,ALICE:2017xrp,Chen:2018tnh,STAR:2019sjh,Ono:2019jxm} and theoretically~\cite{Csernai:1986qf,Dover:1991zn,Scheibl:1998tk,Oh:2009gx,Steinheimer:2012tb,Zhao:2018lyf,Oliinychenko:2018ugs,Vovchenko:2019aoz,Oliinychenko:2020ply,Zhao:2020irc,Sun:2021dlz,Staudenmaier:2021lrg,Oliinychenko:2020znl,Hillmann:2021zgj,Glassel:2021rod,Zhao:2022xkz}. 
It provides important information about the properties of nuclear matter at high densities and temperatures, such as the equation of state~\cite{Sun:2018jhg,Sun:2020pjz,Sun:2020zxy}, the symmetry energy~\cite{Chen:2003qj,Dai:2014rja} and the nucleosynthesis that takes place in stars~\cite{STAR:2011eej,Chen:2018tnh,ALICE:2022zuz}.
Based on coalescence model, it was predicted that the compound yield ratio {\yr} of tritons ($N_t$), deuterons ($N_d$), and protons ($N_p$), is sensitive to the neutron density fluctuations, making it a promising observable to search for the signature of the CP and/or a first-order phase transition in heavy-ion collisions ~\cite{Sun:2017xrx,Sun:2018jhg,Shuryak:2018lgd,Shuryak:2019ikv,Shuryak:2020yrs,Sun:2020pjz,Sun:2020zxy,Sun:2022cxp}. The expected signature of CP is the non-monotonic variation as a function of collision energy. 

 \begin{figure*}[!htpb]
		\centering
		\centerline{\scalebox{0.9}[0.9]{\includegraphics{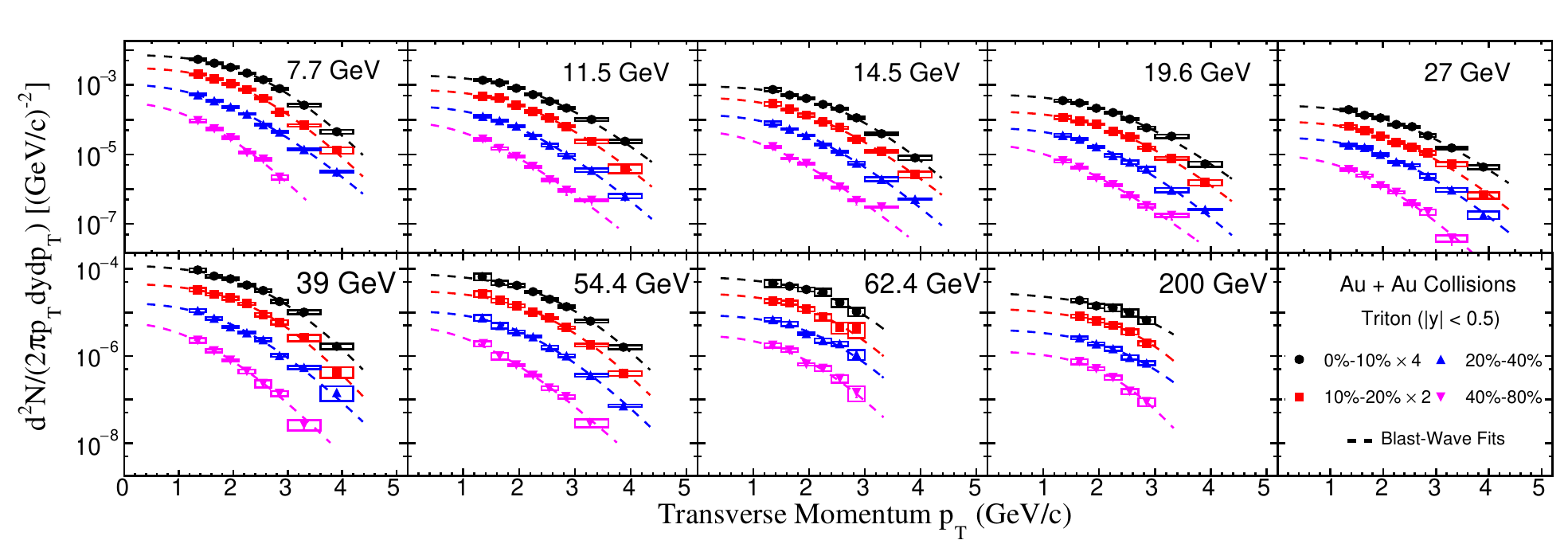}}}
		\caption{\label{f1:ts} Transverse momentum ($p_{T}$) spectra for mid-rapidity ($\left|y\right|$ $<$ 0.5) tritons from 0\%-10\%, 10\%-20\%, 20\%-40\%, and 40\%-80\% centralities in Au+Au collisions at {\snn}= 7.7, 11.5, 14.5, 19.6, 27, 39, 54.4,  62.4, and 200 GeV. Dashed-lines are the corresponding Blast-Wave fits with the profile parameter n = 1. The statistical and systematic uncertainties are shown as vertical lines and boxes, respectively.}
	\end{figure*}

In addition to exploring the QCD phase structure, the systematic measurement of triton yields and yield ratios {\yr} across a broad energy range provide valuable insights into the production mechanism of light nuclei in heavy-ion collisions. Several models have been proposed to explain this production, such as coalescence~\cite{Sato:1981ez, Csernai:1986qf,Oh:2009gx}, thermal~\cite{Mekjian:1978zz,Andronic:2010qu} and dynamical~\cite{Oliinychenko:2018ugs,Vovchenko:2019aoz,Neidig:2021bal}  models. 
In the coalescence model, light nuclei are not considered as point-like particles, but rather have a finite size. Due to the size effect~\cite{Csernai:1986qf}, the coalescence model~\cite{Sun:2018mqq,Zhao:2021dka} predicts that the yield ratio {\yr} should increase as the size of the system or the charged-particle multiplicity decrease. This trend is opposite to what is predicted by thermal model calculations~\cite{Vovchenko:2018fiy}. As a result, the study of the yield ratio can be used to distinguish between these two production mechanisms. The thermal model has been successful in describing the measured yields of hadrons and light (anti-)nuclei in central Pb+Pb collisions at the Large Hadron Collider (LHC)~\cite{Andronic:2017pug,Braun-Munzinger:2018hat}. However, the survival of light nuclei in the hot medium created in heavy-ion collisions remains a puzzle. One possible explanation is that the hadronic re-scatterings play a crucial role during the hadronic expansion phase. Dynamical model calculations with hadronic re-scatterings implemented using both the saha~\cite{Vovchenko:2019aoz} and rate equations~\cite{Neidig:2021bal} show that the deuteron, triton, and helium-3 yields remain unchanged during hadronic expansion. A similar conclusion is obtained in a transport model simulation of hadronic re-scattering processes realized by the dissociation and regeneration of deuterons via the reaction $\pi NN  \leftrightarrow  \pi d$~\cite{Oliinychenko:2018ugs}. Recently, a calculation using the kinetic approach~\cite{Sun:2022xjr} showed that the effects of hadronic re-scatterings during the hadronic expansion stage could reduce the triton and helium-3 yields by approximately a factor of 1.8 from their initial values predicted by the thermal model. The systematic measurement of triton production and the yield ratio {\yr} not only offer a probe into the QCD phase structure, but also serve as valuable experimental evidence for verifying different model calculations and improving our understanding of the production mechanism. 
		
In this letter, we report triton production at mid-rapidity ($|y|<0.5$) in Au+Au collisions at {\snn}= 7.7, 11.5, 14.5, 19.6, 27, 39, 54.4, 62.4, and 200 GeV measured by the STAR experiment from the first phase of the Beam Energy Scan (BES-I, 2010-2017) program at RHIC~\cite{STAR:2002eio}. The results presented are analyzed from minimum bias events of Au+Au collisions, occurring within +/-30 cm for 200 GeV and +/-40 cm for other energies of the nominal interaction point along the beam axis. Collision centralities are determined by fitting the measured charged particle multiplicities within pseudorapidity $|\eta|<0.5$ with a Monte Carlo Glauber model~\cite{STAR:2008med}. The selected tracks are required to have a distance of closest approach (DCA) to the primary collision vertex of less than 1 cm and have at least 20 hit points measured in the Time Projection Chamber (TPC). Triton identification is performed using information from the TPC and Time-Of-Flight (TOF) detectors~\cite{Llope:2005yw}. Based on the measurement of the specific ionization energy deposited ($dE/dx$) by charged particles in the TPC, a new variable $z$ is defined to properly deconvolve these effects into a Gaussian. It is defined as
$z=\text{ln}\left(\frac{\langle dE/dx\rangle}{\langle dE/dx\rangle_\text{B}}\right)$, 
where $\langle dE/dx\rangle_\text{B}$ is the Bichsel function for each particle species. A cut of $|z|\leqslant 0.3$ is applied to remove most contamination from the triton raw signals. To extract the raw triton yields, the mass squared ($m^2$) distributions from the TOF detector were used, which is defined as
$m^2=p^2\left(\frac{c^2t^2}{L^2}-1\right)$, 
where $t$, $L$, and $c$ are the particle flight time, track length, and speed of light, respectively. The $m^2$ distribution is fit with a superposition of a Gaussian function and an exponential tail for the triton signal and background, respectively. 

The final triton $p_{T}$ spectra are obtained by applying several corrections to the raw spectra, including corrections for the tracking efficiency, low momentum energy loss, and absorption of light nuclei by the detector material. These corrections were calculated using the embedding simulations from the experiment~\cite{STAR:2019sjh,STAR:2001pbk}. Because the TOF detector is used to identify tritons at high $p_{T}$, we also need to correct for the TOF matching efficiency, defined as the ratio of the number of tracks matched in the TOF to the number of total tracks in the TPC within the same acceptance. The point-to-point systematic uncertainties on the spectra are estimated by varying track selection, analysis cuts and by assessing the sample purity from the $dE/dx$ measurement. Track selection and particle identification contribute by $\sim$3\% and signal extraction contributes by less than $\sim$2\% at low $p_{T}$ and increasing to $\sim$10\% at high $p_{T}$ due to the reduced resolution of the TPC. A correlated systematic uncertainty of 5\% is estimated for all spectra and is dominated by uncertainties in the Monte Carlo determination of reconstruction efficiencies. All of these uncertainties are added in quadrature to obtain the final systematic uncertainties.

Figure~\ref{f1:ts} shows the $p_{T}$ spectra of identified tritons measured at mid-rapidity ($|y|<0.5$) in Au+Au collisions at {\snn}= 7.7, 11.5, 14.5, 19.6, 27, 39, 54.4, 62.4, and 200 GeV for 0\%-10\%, 10\%-20\%, 20\%-40\%, and 40\%-80\% centralities. 
The $p_T$-integrated particle yields ($dN/dy$) are calculated from the measured $p_T$ range and extrapolated to the unmeasured regions with individual Blast-Wave model fits~\cite{Schnedermann:1993ws}. The extrapolation of the $p_T$ spectra to the unmeasured low $p_{T}$ range is the main source of systematic uncertainty on $dN/dy$, which is estimated by fitting the $p_T$ spectra with different functions and comparing the extrapolated values. The systematic uncertainty of yield extrapolations is estimated to be around 5\%-20\%. All of the mid-rapidity proton $p_{T}$ spectra and $dN/dy$ in Au+Au collisions at RHIC energies presented in this paper have been corrected for the weak decay feed-down via a data-driven approach~\cite{STAR:2006uve}, which uses the inclusive proton spectra~\cite{STAR:2008med, STAR:2017sal} and the yields of strange hadrons measured by the STAR experiment~\cite{STAR:2019bjj,supp:ref}. In a previously published STAR paper~\cite{STAR:2017ieb}, the proton feed-down correction was done by using a UrQMD + GEANT simulation, which underestimates the proton feed-down contributions from weak decays. 
\begin{figure}[!htp]
		\centering
		\centerline{\scalebox{0.4}[0.4]{\includegraphics{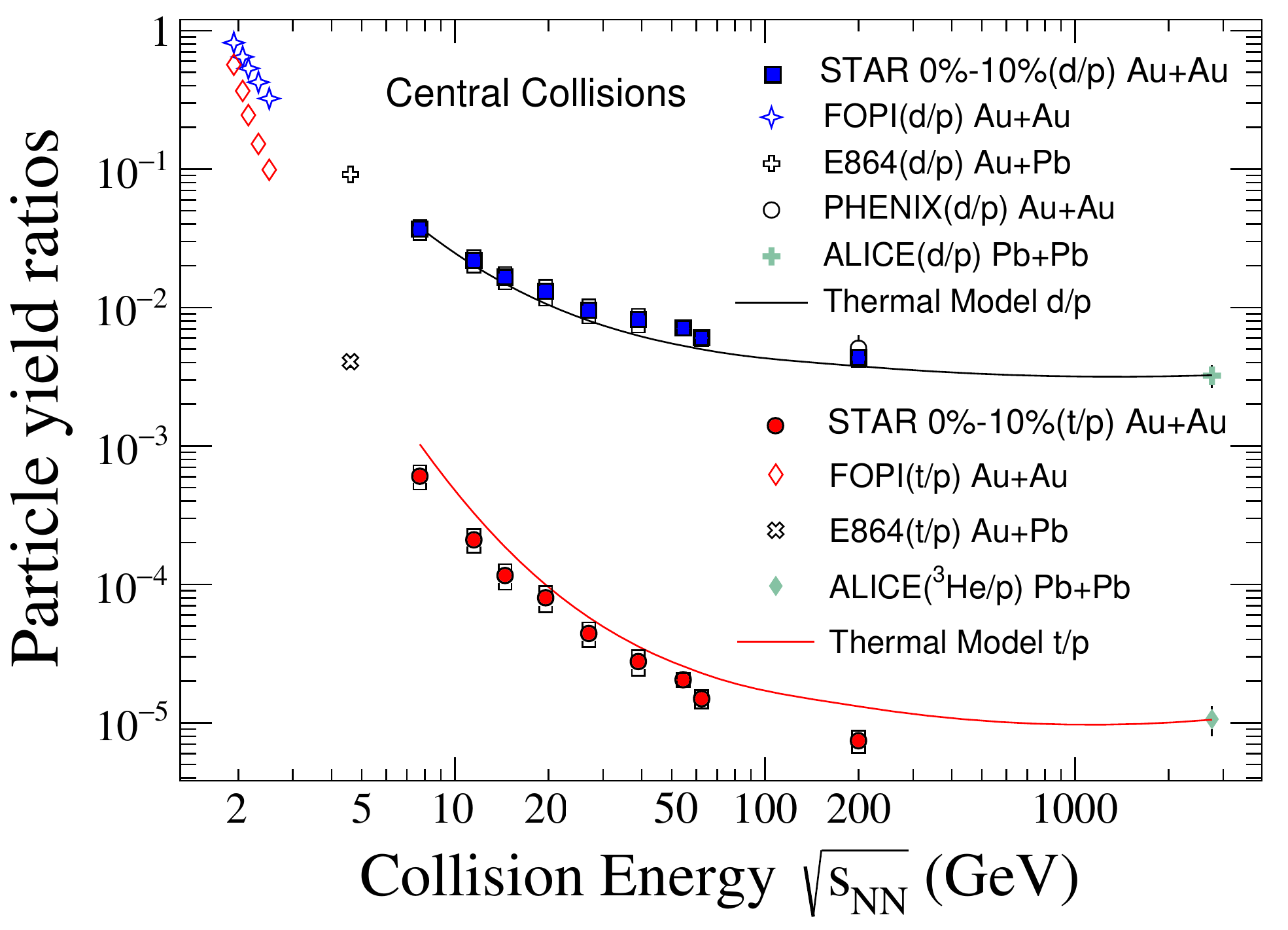}}}
		\caption{\label{f2:yr} Collision energy dependence of the mid-rapidity ratios {\dop} (blue solid squares)  and {\tp} (red solid circles) from the top 0\%-10\% central Au+Au collisions. Statistical and systematic uncertainties are shown as vertical lines and brackets, respectively. For comparison, results from FOPI~\cite{FOPI:2006ifg}, E864~\cite{E864:2000auv}, 
		PHENIX~\cite{PHENIX:2004vqi, PHENIX:2003iij}, and ALICE~\cite{ALICE:2015wav} are also shown. The lines are results from the thermal model using chemical freeze-out conditions from Ref.~\cite{Vovchenko:2015idt,Vovchenko:2020dmv}}
\end{figure}

Figure~\ref{f2:yr} shows the energy dependence of $dN/dy$ ratios, {\dop}~\cite{STAR:2019sjh} and {\tp}, in the mid-rapidity of central heavy-ion collisions from different experiments, including the  FOPI~\cite{FOPI:2006ifg}, E864~\cite{E864:2000auv}, PHENIX~\cite{PHENIX:2004vqi, PHENIX:2003iij}, and ALICE~\cite{ALICE:2015wav} experiments. 
Both the \tp\ and \dop\ ratios decrease monotonically with increasing collision energy and the differences between the ratios get smaller at lower collision energies. The solid lines represent the results calculated from the thermal model which does not include excited nuclei~\cite{Vovchenko:2019pjl}, in which the parametrization of chemical freeze-out temperature and $\mu_{B}$ from Ref.~\cite{Vovchenko:2015idt,Vovchenko:2020dmv} are used. Quantitatively, the thermal model describes the \dop\ ratios well, but it systematically overestimates the \tp\ ratios except for the results from central Pb+Pb collisions at {\snn}= 2.76 TeV~\cite{ALICE:2015wav}. In addition, the coalescence model, which predicts light nuclei production at mid-rapidity based on baryon density ($\rho_{B}$) via the relationship $N_{A}/N_{p} \propto \rho_{B}^{A-1}$, can also describe energy dependence trends~\cite{Zhao:2021dka}. 
\begin{figure}[!htp]
		\centering
		\centerline{\scalebox{0.42}[0.42]{\includegraphics{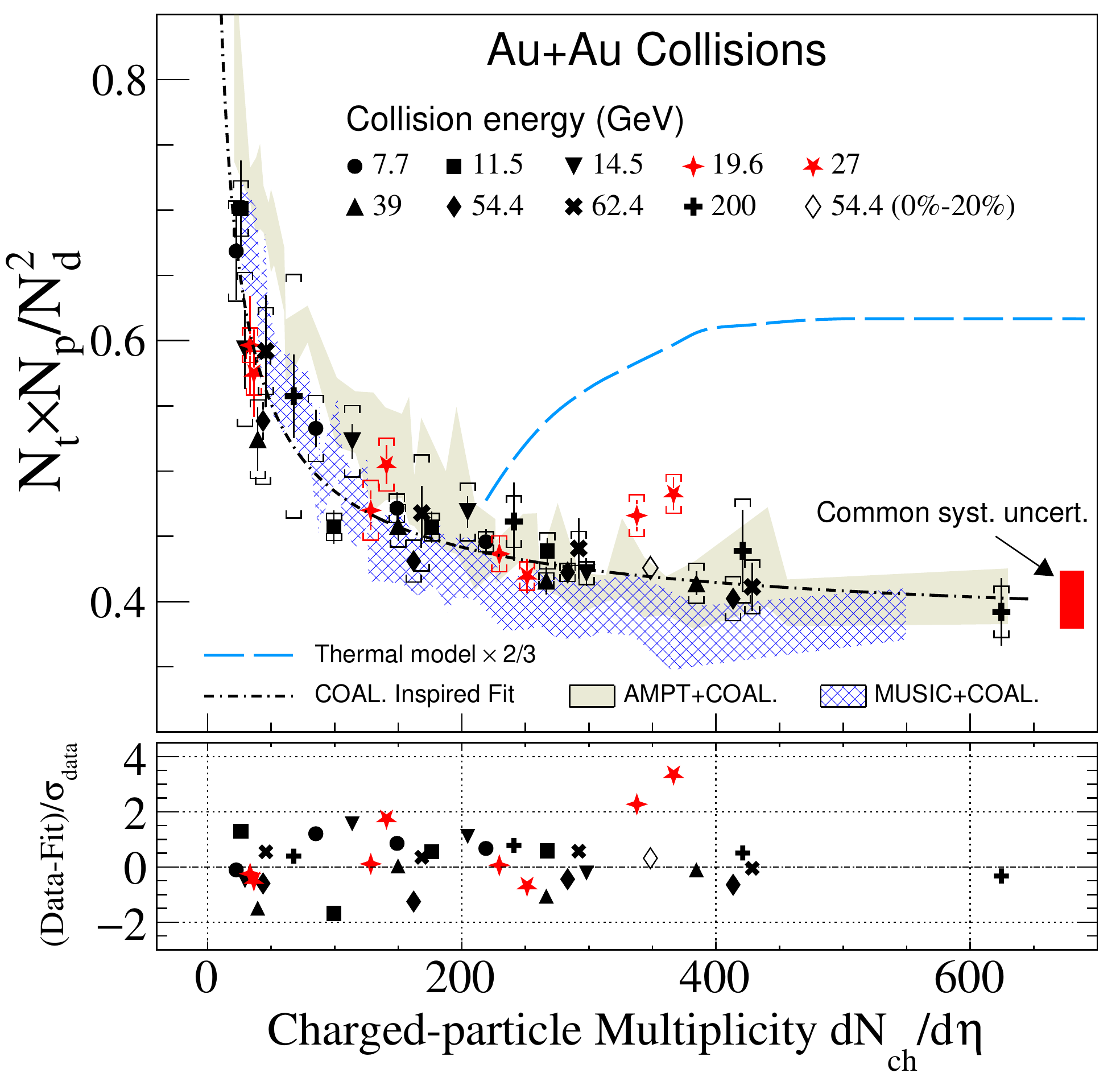}}}
		\caption{\label{f3:ratio} The yield ratio {\yr} as a function of charged-particle multiplicity {\dndeta} ($|\eta|<0.5$) in Au+Au collisions at {\snn}= 7.7 -- 200 GeV for 0\%-10\%, 10\%-20\%, 20\%-40\%, and 40\%-80\% centralities. Statistical and systematic uncertainties are shown as vertical lines and brackets, respectively. The black dot-dashed line denotes the coalescence-inspired fit. The open diamond denotes the yield ratio of 0\%-20\% central Au+Au collisions at {\snn}= 54.4 GeV. The red shaded vertical band on the right side of the figure represents the multiplicity independent systematic uncertainties on these ratios. The significance of the deviation relative to the fit is shown in the lower panel. The results calculated from thermal model are shown as the blue long-dashed line. Calculations from AMPT and MUSIC+UrQMD hybrid models~\cite{Sun:2018mqq,Zhao:2021dka} are shown as shaded bands.}
	\end{figure}	
\begin{figure*}[!htp]
		\centering
		\centerline{\scalebox{0.7}[0.7]{\includegraphics{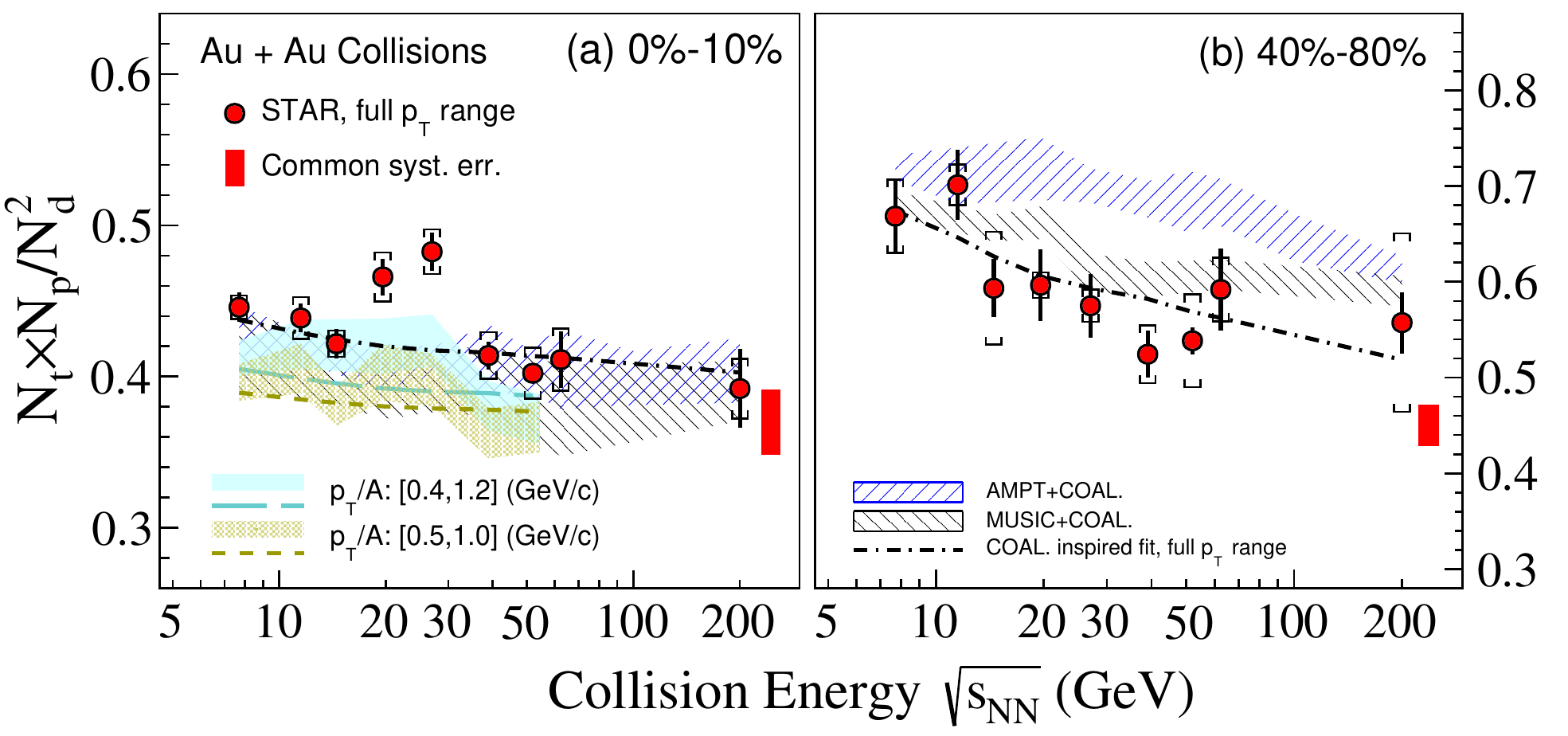}}} \caption{\label{f4:yrerg} Collision energy, centrality, and $p_{T}$ dependence of the yield ratio {\yr} in Au+Au collisions at RHIC. Solid circles are the results from 0\%-10\% central (left panel) and 40\%-80\% peripheral (right panel) collisions. Colored-bands in panel (a) denote $p_{T}$ acceptance dependence, for which the statistical and systematic uncertainties are added in quadrature. Red solid circles are the final results with extrapolation to the full $p_{T}$ range. Statistical and systematic uncertainties are shown as bars and brackets, respectively.
		Red vertical bands on the right side of panels represent the common systematic uncertainties. Dashed lines are the coalescence baselines obtained from the coalescence-inspired fit. Shaded areas denote the calculations from hadronic transport AMPT and MUSIC+UrQMD hybrid models~\cite{Zhao:2021dka}.}
	\end{figure*}	

As mentioned earlier, the yield ratio {\yr} is predicted to be sensitive to the local baryon density fluctuations and can be used to probe the QCD phase structure. Figure~\ref{f3:ratio} shows the charged-particle multiplicity {\dndeta} ($|\eta|<0.5$) dependence of the yield ratio {\yr} in Au+Au collisions at {\snn}= 7.7 - 200 GeV. The data from each collision energy presented in the figure include four centrality bins: 0\%-10\%, 10\%-20\%, 20\%-40\%, and 40\%-80\%, in addition, a single 0\%-20\% centrality bin is also presented for 54.4 GeV. It is observed that the yield ratio {\yr} exhibits scaling, regardless of collision energy and centrality. The shaded bands in Fig.~\ref{f3:ratio} are the corresponding results from the calculations of hadronic transport AMPT and MUSIC+UrQMD hybrid models~\cite{Zhao:2021dka}. MUSIC is a (3+1)D viscous hydrodynamics model~\cite{Shen:2017bsr,Denicol:2018wdp}, which conserves both energy-momentum and baryon number and is used to describe the dynamical evolution of the QGP. To provide a reliable baseline, neither critical point nor first-order phase transition is included in the AMPT and MUSIC+UrQMD hybrid model calculations. These two models are employed to generate the nucleon phase space at kinetic freeze-out, when light nuclei are formed via nucleon coalescence. It is found that the overall trend of the experimental data is well described by the model calculations. The light blue dashed line is the result calculated from the thermal model at chemical freeze-out~\cite{Vovchenko:2015idt,Vovchenko:2020dmv} for central Au+Au collisions, which overestimates the experimental data by more than a factor of two at {\dndeta} $\sim 600$. As discussed in Ref.~\cite{Sun:2022xjr}, this overestimation could be due to the effects of hadronic re-scatterings during hadronic expansion, which reduce the triton and helium-3 yields by about a factor of 1.8 from their initial values predicted by thermal model. However, this cannot explain the agreement between the thermal model calculations and the $\mathrm{N}_{^{3}\mathrm{He}} \times \mathrm{N}_p/\mathrm{N}_d^2$ ratio from central Pb+Pb collisions at {\snn}= 2.76 TeV where {\dndeta} $\sim 1100$~\cite{ALICE:2015wav, Oliinychenko:2020ply}. Obviously, further investigations are needed to understand the discrepancy.

 The black dot-dashed line is a fit to the data based on the coalescence model. As discussed in Ref.~\cite{Zhao:2021dka}, assuming a thermal equilibrated and static spherical Gaussian nucleon source, one can obtain the fit function as:
\begin{equation}
    \frac{\mathrm{N}_t \times \mathrm{N}_p}{\mathrm{N}_d^2} =  p_0\times \left (\frac{R^{2} + \frac{2}{3}r_{d}^{2}     }{R^{2} + \frac{1}{2}r_{t}^{2}} \right)^{3},
    \label{eq:size}
\end{equation}
where $R = p_{1}\times (dN_{ch}/d\eta)^{1/3}$ denotes the radius of the spherical nucleon emission source. $r_{d}$ = 1.96 fm and $r_{t}$ = 1.59 fm are the nucleonic point root-mean-square radius of deuteron and triton~\cite{Ropke:2008qk}, respectively. $p_{0}$ and $p_{1}$ are the two fitting parameters where the best fit values are 0.37 $\pm$ 0.008 and 0.75 $\pm$ 0.04, respectively. At small values of {\dndeta}, when the system size is comparable to the size of light nuclei, the yield ratio shows a rapid increase with decreasing {\dndeta}, while it saturates at large charged-particle multiplicity.
The general trend of the yield ratio {\yr} is driven by the interplay between the finite size of light nuclei and the overall size of the fireball created in heavy-ion collisions. This provides strong evidence that nucleon coalescence is the correct formation mechanism to describe the light nuclei production in such collisions. If we use the coalescence-inspired fit as the baseline, the lower panel of the Fig.~\ref{f3:ratio} shows that most of the measurements are within significance of 2$\sigma$ from the coalescence baseline, except there are enhancements observed for the yield ratios in the 0\%-10\% most central Au+Au collisions at {\snn}= 19.6 and 27 GeV with significance of 2.3$\sigma$ and 3.4$\sigma$, respectively, and for a combined significance of 4.1$\sigma$, as shown in the lower panel of Fig.~\ref{f3:ratio}. The yield ratio of 0\%-20\% central Au+Au collisions at 54.4 GeV is also shown in Fig.~\ref{f3:ratio} as an open diamond. It agrees with the coalescence baseline at the same value of {\dndeta} as those data points from central collisions at {\snn}= 19.6 and 27 GeV. Therefore, the observed enhancement may be driven by the baryon density rather than the overall size of the system which is proportional to the charged-particle density  {\dndeta}. In order to understand the origin of the observed enhancement in the ratios, further dynamical modeling of heavy-ion collisions with a realistic equation of state is needed.

Figure~\ref{f4:yrerg} shows the energy dependence of the yield ratio {\yr} at mid-rapidity in central (0\%-10\%) and peripheral (40\%-80\%) Au+Au collisions at {\snn}= 7.7 - 200 GeV. 
For comparison, the coalescence baselines obtained by fitting the {\dndeta} dependence of the yield ratio as shown in Fig. ~\ref{f3:ratio} and the calculations of AMPT, MUSIC+UrQMD hybrid models are displayed in Fig.~\ref{f4:yrerg}. For the 0\%-10\% most central Au+Au collisions, the yield ratios are consistent with the coalescence baseline and model calculations, except for the enhancements of the yield ratios to coalescence baseline with a significance of 2.3$\sigma$ and 3.4$\sigma$ observed at {\snn}= 19.6 and 27 GeV, respectively. 
The colored bands in panel (a) denote the yield ratios, in which the proton, deuteron, and triton yields are obtained from the commonly measured $p_{T}/A$ range without any extrapolation. The enhancements and the significance of the measurements decrease with smaller $p_{T}$ acceptance in the region of interest. The combined (19.6 and 27 GeV) significance of enhancements to the corresponding coalescence baselines for 0.5 $\leq$ $p_{T}/A$ $\leq$ 1.0 GeV/c, 0.4 $\leq$ $p_{T}/A$ $\leq$ 1.2 GeV/c, and the full $p_{T}/A$ range are 1.6$\sigma$, 2.5$\sigma$, and 4.1$\sigma$, respectively. In the model calculations, the physics of the critical point or first-order phase transition are not included. Therefore, the non-monotonic behavior observed in the energy dependence of the yield ratio {\yr} from 0\%-10\% central Au+Au collisions may be due to the enhanced baryon density fluctuations induced by the critical point or first-order phase transition in heavy-ion collisions. The right panel of Fig.~\ref{f4:yrerg} shows the energy dependence of the yield ratio in peripheral (40\%-80\%) Au+Au collisions. Within uncertainties, the experimental data can be well described by the coalescence baseline (black-dashed line) whereas the calculations from AMPT and MUSIC+UrQMD hybrid models overestimate the data. 

In summary, we present the triton production and the yield ratio {\yr} in mid-rapidity Au+Au collisions at {\snn}= 7.7 -- 200 GeV measured by the STAR experiment at RHIC. The yield ratio {\yr} shows a monotonic decrease with increasing charged-particle multiplicity ({\dndeta}) and exhibits a scaling behavior, which can be attributed to the formation of deuteron and triton via nucleon coalescence. The thermal model, however, overestimates the triton over proton yield ratio {\tp} and the {\yr} ratio at RHIC energies, possibly due to the effect of hadronic re-scatterings during the hadronic expansion stage. In the 0\%-10\% most central Au+Au collisions at {\snn}= 19.6 and 27 GeV, {\yr} shows enhancements relative to the coalescence baseline with a significance of 2.3$\sigma$ and 3.4$\sigma$, respectively, and a combined significance of 4.1$\sigma$. The significance of the measurement decreases with reduced $p_{T}$ range, indicating that the possible enhancement may have a strong dependence on the $p_{T}$ acceptance. In peripheral collisions, similar to data, model calculations have a smooth decreasing trend as a function of energy. Further studies from dynamical modeling of heavy-ion collisions with a realistic equation of state are required to confirm if the enhancements are due to large baryon density fluctuations near the critical point. These systematic measurements of triton yields and yield ratios over a broad energy range provide important insights into the production dynamics of light nuclei and our understanding of the QCD phase diagram.



We thank Drs. L. W. Chen, C.~M. Ko, V. Koch, D. Oliinychenko, J. Steinheimer, K. J. Sun, V. Vovchenko and W. Zhao for interesting discussions about light nuclei production in heavy-ion collisions. 
We thank the RHIC Operations Group and RCF at BNL, the NERSC Center at LBNL, and the Open Science Grid consortium for providing resources and support.  This work was supported in part by the Office of Nuclear Physics within the U.S. DOE Office of Science, the U.S. National Science Foundation, National Natural Science Foundation of China, Chinese Academy of Science, the Ministry of Science and Technology of China and the Chinese Ministry of Education, the Higher Education Sprout Project by Ministry of Education at NCKU, the National Research Foundation of Korea, Czech Science Foundation and Ministry of Education, Youth and Sports of the Czech Republic, Hungarian National Research, Development and Innovation Office, New National Excellency Programme of the Hungarian Ministry of Human Capacities, Department of Atomic Energy and Department of Science and Technology of the Government of India, the National Science Centre and WUT ID-UB of Poland, the Ministry of Science, Education and Sports of the Republic of Croatia, German Bundesministerium f\"ur Bildung, Wissenschaft, Forschung and Technologie (BMBF), Helmholtz Association, Ministry of Education, Culture, Sports, Science, and Technology (MEXT) and Japan Society for the Promotion of Science (JSPS).


\FloatBarrier


\begin{thebibliography}{10}
\bibitem{Rajagopal:2000wf}
K.~Rajagopal and F.~Wilczek,
[arXiv:hep-ph/0011333 [hep-ph]].

\bibitem{STAR:2010vob}
M.~M.~Aggarwal \textit{et al.} [STAR],
[arXiv:1007.2613 [nucl-ex]].

\bibitem{Aoki:2006we}
Y.~Aoki, G.~Endrodi, Z.~Fodor, S.~D.~Katz and K.~K.~Szabo,
Nature \textbf{443}, 675-678 (2006)

\bibitem{HotQCD:2018pds}
A.~Bazavov \textit{et al.} [HotQCD],
Phys. Lett. B \textbf{795}, 15-21 (2019)

\bibitem{Ejiri:2008xt}
S.~Ejiri,
Phys. Rev. D \textbf{78}, 074507 (2008)

\bibitem{Fischer:2018sdj}
C.~S.~Fischer,
Prog. Part. Nucl. Phys. \textbf{105}, 1-60 (2019)

\bibitem{Fu:2019hdw}
W.~j.~Fu, J.~M.~Pawlowski and F.~Rennecke,
Phys. Rev. D \textbf{101}, no.5, 054032 (2020)

\bibitem{Gao:2020fbl}
F.~Gao and J.~M.~Pawlowski,
Phys. Lett. B \textbf{820}, 136584 (2021)

\bibitem{Halasz:1998qr}
A.~M.~Halasz, A.~D.~Jackson, R.~E.~Shrock, M.~A.~Stephanov and J.~J.~M.~Verbaarschot,
Phys. Rev. D \textbf{58}, 096007 (1998)

\bibitem{Stephanov:2004wx}
M.~A.~Stephanov,
Prog. Theor. Phys. Suppl. \textbf{153}, 139-156 (2004)

\bibitem{Fukushima:2010bq}
K.~Fukushima and T.~Hatsuda,
Rept. Prog. Phys. \textbf{74}, 014001 (2011)

\bibitem{Gupta:2011wh}
S.~Gupta, X.~Luo, B.~Mohanty, H.~G.~Ritter and N.~Xu,
Science \textbf{332}, 1525-1528 (2011)

\bibitem{STAR:2013gus}
L.~Adamczyk \textit{et al.} [STAR],
Phys. Rev. Lett. \textbf{112}, 032302 (2014)

\bibitem{Bzdak:2019pkr}
A.~Bzdak, S.~Esumi, V.~Koch, J.~Liao, M.~Stephanov and N.~Xu,
Phys. Rept. \textbf{853}, 1-87 (2020)

\bibitem{Luo:2017faz}
X.~Luo and N.~Xu,
Nucl. Sci. Tech. \textbf{28}, no.8, 112 (2017)

\bibitem{STAR:2020tga}
J.~Adam \textit{et al.} [STAR],
Phys. Rev. Lett. \textbf{126}, no.9, 092301 (2021)

\bibitem{STAR:2021iop}
M.~Abdallah \textit{et al.} [STAR],
Phys. Rev. C \textbf{104}, no.2, 024902 (2021)

\bibitem{STAR:2014shf}
L.~Adamczyk \textit{et al.} [STAR],
Phys. Rev. C \textbf{92}, no.1, 014904 (2015)

\bibitem{STAR:2020dav}
J.~Adam \textit{et al.} [STAR],
Phys. Rev. C \textbf{103}, no.3, 034908 (2021)

\bibitem{STAR:2014clz}
L.~Adamczyk \textit{et al.} [STAR],
Phys. Rev. Lett. \textbf{112}, no.16, 162301 (2014)

\bibitem{STAR:2017okv}
L.~Adamczyk \textit{et al.} [STAR],
Phys. Rev. Lett. \textbf{120}, no.6, 062301 (2018)

\bibitem{STAR:2023jpm}
 [STAR],
[arXiv:2301.11062 [nucl-ex]].




\bibitem{Cocconi:1960zz}
V.~T.~Cocconi, T.~Fazzini, G.~Fidecaro, M.~Legros, N.~H.~Lipman and A.~W.~Merrison,
Phys. Rev. Lett. \textbf{5}, 19-21 (1960)

\bibitem{E814:1994kon}
J.~Barrette \textit{et al.} [E814],
Phys. Rev. C \textbf{50}, 1077-1084 (1994)

\bibitem{E864:2000auv}
T.~A.~Armstrong \textit{et al.} [E864],
Phys. Rev. C \textbf{61}, 064908 (2000)

\bibitem{Albergo:2002gi}
S.~Albergo, R.~Bellwied, M.~Bennett, B.~Bonner, H.~Caines, W.~Christie, S.~Costa, H.~J.~Crawford, M.~Cronqvist and R.~Debbe, \textit{et al.}
Phys. Rev. C \textbf{65}, 034907 (2002)


\bibitem{FOPI:2010xrt}
W.~Reisdorf \textit{et al.} [FOPI],
Nucl. Phys. A \textbf{848}, 366-427 (2010)

\bibitem{ALICE:2015wav}
J.~Adam \textit{et al.} [ALICE],
Phys. Rev. C \textbf{93}, no.2, 024917 (2016)

\bibitem{NA49:2016qvu}
T.~Anticic \textit{et al.} [NA49],
Phys. Rev. C \textbf{94}, no.4, 044906 (2016)

\bibitem{STAR:2016ydv}
L.~Adamczyk \textit{et al.} [STAR],
Phys. Rev. C \textbf{94}, no.3, 034908 (2016)


\bibitem{ALICE:2017xrp}
S.~Acharya \textit{et al.} [ALICE],
Phys. Rev. C \textbf{97}, no.2, 024615 (2018)


\bibitem{Chen:2018tnh}
J.~Chen, D.~Keane, Y.~G.~Ma, A.~Tang and Z.~Xu,
Phys. Rept. \textbf{760}, 1-39 (2018)

\bibitem{STAR:2019sjh}
J.~Adam \textit{et al.} [STAR],
Phys. Rev. C \textbf{99}, no.6, 064905 (2019)

\bibitem{Ono:2019jxm}
A.~Ono,
Prog. Part. Nucl. Phys. \textbf{105}, 139-179 (2019)

\bibitem{Csernai:1986qf}
L.~P.~Csernai and J.~I.~Kapusta,
Phys. Rept. \textbf{131}, 223-318 (1986)

\bibitem{Dover:1991zn}
C.~B.~Dover, U.~W.~Heinz, E.~Schnedermann and J.~Zimanyi,
Phys. Rev. C \textbf{44}, 1636-1654 (1991)

\bibitem{Scheibl:1998tk}
R.~Scheibl and U.~W.~Heinz,
Phys. Rev. C \textbf{59}, 1585-1602 (1999)

\bibitem{Oh:2009gx}
Y.~Oh, Z.~W.~Lin and C.~M.~Ko,
Phys. Rev. C \textbf{80}, 064902 (2009)

\bibitem{Steinheimer:2012tb}
J.~Steinheimer, K.~Gudima, A.~Botvina, I.~Mishustin, M.~Bleicher and H.~Stocker,
Phys. Lett. B \textbf{714}, 85-91 (2012)

\bibitem{Zhao:2018lyf}
W.~Zhao, L.~Zhu, H.~Zheng, C.~M.~Ko and H.~Song,
Phys. Rev. C \textbf{98}, no.5, 054905 (2018)

\bibitem{Oliinychenko:2018ugs}
D.~Oliinychenko, L.~G.~Pang, H.~Elfner and V.~Koch,
Phys. Rev. C \textbf{99}, no.4, 044907 (2019)

\bibitem{Vovchenko:2019aoz}
V.~Vovchenko, K.~Gallmeister, J.~Schaffner-Bielich and C.~Greiner,
Phys. Lett. B \textbf{800}, 135131 (2020)

\bibitem{Oliinychenko:2020ply}
D.~Oliinychenko,
Nucl. Phys. A \textbf{1005}, 121754 (2021)

\bibitem{Zhao:2020irc}
W.~Zhao, C.~Shen, C.~M.~Ko, Q.~Liu and H.~Song,
Phys. Rev. C \textbf{102}, no.4, 044912 (2020)

\bibitem{Sun:2021dlz}
K.~J.~Sun, R.~Wang, C.~M.~Ko, Y.~G.~Ma and C.~Shen,

\bibitem{Staudenmaier:2021lrg}
J.~Staudenmaier, D.~Oliinychenko, J.~M.~Torres-Rincon and H.~Elfner,
Phys. Rev. C \textbf{104}, no.3, 034908 (2021)

\bibitem{Oliinychenko:2020znl}
D.~Oliinychenko, C.~Shen and V.~Koch,
Phys. Rev. C \textbf{103}, no.3, 034913 (2021)

\bibitem{Hillmann:2021zgj}
P.~Hillmann, K.~K\"afer, J.~Steinheimer, V.~Vovchenko and M.~Bleicher,
J. Phys. G \textbf{49}, no.5, 055107 (2022)

\bibitem{Glassel:2021rod}
S.~Gl\"a\ss{}el, V.~Kireyeu, V.~Voronyuk, J.~Aichelin, C.~Blume, E.~Bratkovskaya, G.~Coci, V.~Kolesnikov and M.~Winn,
Phys. Rev. C \textbf{105}, no.1, 014908 (2022)

\bibitem{Zhao:2022xkz}
X.~Y.~Zhao, Y.~T.~Feng, F.~L.~Shao, R.~Q.~Wang and J.~Song,
Phys. Rev. C \textbf{105}, 054908 (2022)

\bibitem{Sun:2018jhg}
K.~J.~Sun, L.~W.~Chen, C.~M.~Ko, J.~Pu and Z.~Xu,
Phys. Lett. B \textbf{781}, 499-504 (2018)

\bibitem{Sun:2020pjz}
K.~J.~Sun, C.~M.~Ko, F.~Li, J.~Xu and L.~W.~Chen,
Eur. Phys. J. A \textbf{57}, no.11, 313 (2021)

\bibitem{Sun:2020zxy}
K.~J.~Sun, F.~Li and C.~M.~Ko,
Phys. Lett. B \textbf{816}, 136258 (2021)

\bibitem{Chen:2003qj}
L.~W.~Chen, C.~M.~Ko and B.~A.~Li,
Phys. Rev. C \textbf{68}, 017601 (2003)

\bibitem{Dai:2014rja}
Z.~T.~Dai, D.~Q.~Fang, Y.~G.~Ma, X.~G.~Cao and G.~Q.~Zhang,
Phys. Rev. C \textbf{89}, no.1, 014613 (2014)

\bibitem{STAR:2011eej}
H.~Agakishiev \textit{et al.} [STAR],
Nature \textbf{473}, 353 (2011)
[erratum: Nature \textbf{475}, 412 (2011)]

\bibitem{ALICE:2022zuz}
S.~Acharya \textit{et al.} [ALICE],
Nature Phys. \textbf{19}, no.1, 61-71 (2023)

\bibitem{Sun:2017xrx}
K.~J.~Sun, L.~W.~Chen, C.~M.~Ko and Z.~Xu,
Phys. Lett. B \textbf{774}, 103-107 (2017)

\bibitem{Shuryak:2018lgd}
E.~Shuryak and J.~M.~Torres-Rincon,
Phys. Rev. C \textbf{100}, no.2, 024903 (2019)

\bibitem{Shuryak:2019ikv}
E.~Shuryak and J.~M.~Torres-Rincon,
Phys. Rev. C \textbf{101}, no.3, 034914 (2020)

\bibitem{Shuryak:2020yrs}
E.~Shuryak and J.~M.~Torres-Rincon,
Eur. Phys. J. A \textbf{56}, no.9, 241 (2020)

\bibitem{Sun:2022cxp}
K.~J.~Sun, W.~H.~Zhou, L.~W.~Chen, C.~M.~Ko, F.~Li, R.~Wang and J.~Xu,
[arXiv:2205.11010 [nucl-th]].

\bibitem{Sato:1981ez}
H.~Sato and K.~Yazaki,
Phys. Lett. B \textbf{98}, 153-157 (1981)



\bibitem{Mekjian:1978zz}
A.~Z.~Mekjian,
Phys. Rev. C \textbf{17}, 1051-1070 (1978)


\bibitem{Andronic:2010qu}
A.~Andronic, P.~Braun-Munzinger, J.~Stachel and H.~Stocker,
Phys. Lett. B \textbf{697}, 203-207 (2011)

\bibitem{Neidig:2021bal}
T.~Neidig, K.~Gallmeister, C.~Greiner, M.~Bleicher and V.~Vovchenko,
Phys. Lett. B \textbf{827}, 136891 (2022)

\bibitem{Sun:2018mqq}
K.~J.~Sun, C.~M.~Ko and B.~D\"onigus,
Phys. Lett. B \textbf{792}, 132-137 (2019)

\bibitem{Zhao:2021dka}
W.~Zhao, K.~j.~Sun, C.~M.~Ko and X.~Luo,
Phys. Lett. B \textbf{820}, 136571 (2021)

\bibitem{Vovchenko:2018fiy}
V.~Vovchenko, B.~D\"onigus and H.~Stoecker,
Phys. Lett. B \textbf{785}, 171-174 (2018)








\bibitem{Andronic:2017pug}
A.~Andronic, P.~Braun-Munzinger, K.~Redlich and J.~Stachel,
Nature \textbf{561}, no.7723, 321-330 (2018)

\bibitem{Braun-Munzinger:2018hat}
P.~Braun-Munzinger and B.~D\"onigus,
Nucl. Phys. A \textbf{987}, 144-201 (2019)


\bibitem{Sun:2022xjr}
K.~J.~Sun, R.~Wang, C.~M.~Ko, Y.~G.~Ma and C.~Shen,
[arXiv:2207.12532 [nucl-th]].


\bibitem{STAR:2002eio}
K.~H.~Ackermann \textit{et al.} [STAR],
Nucl. Instrum. Meth. A \textbf{499}, 624-632 (2003)

\bibitem{STAR:2008med}
B.~I.~Abelev \textit{et al.} [STAR],
Phys. Rev. C \textbf{79}, 034909 (2009)

\bibitem{Llope:2005yw}
W.~J.~Llope,
Nucl. Instrum. Meth. B \textbf{241}, 306-310 (2005)

\bibitem{STAR:2001pbk}
C.~Adler \textit{et al.} [STAR],
Phys. Rev. Lett. \textbf{87}, 262301 (2001)
[erratum: Phys. Rev. Lett. \textbf{87}, 279902 (2001)]

\bibitem{FOPI:2006ifg}
W.~Reisdorf \textit{et al.} [FOPI],
Nucl. Phys. A \textbf{781}, 459-508 (2007)

\bibitem{PHENIX:2004vqi}
S.~S.~Adler \textit{et al.} [PHENIX],
Phys. Rev. Lett. \textbf{94}, 122302 (2005)


\bibitem{PHENIX:2003iij}
S.~S.~Adler \textit{et al.} [PHENIX],
Phys. Rev. C \textbf{69}, 034909 (2004)

\bibitem{Vovchenko:2015idt}
V.~Vovchenko, V.~V.~Begun and M.~I.~Gorenstein,
Phys. Rev. C \textbf{93}, no.6, 064906 (2016)

\bibitem{Vovchenko:2020dmv}
V.~Vovchenko, B.~D\"onigus, B.~Kardan, M.~Lorenz and H.~Stoecker,
Phys. Lett. \textbf{B}, 135746 (2020)

\bibitem{Schnedermann:1993ws}
E.~Schnedermann, J.~Sollfrank and U.~W.~Heinz,
Phys. Rev. C \textbf{48}, 2462-2475 (1993)

\bibitem{STAR:2006uve}
B.~I.~Abelev \textit{et al.} [STAR],
Phys. Rev. Lett. \textbf{97}, 152301 (2006)

\bibitem{STAR:2017sal}
L.~Adamczyk \textit{et al.} [STAR],
Phys. Rev. C \textbf{96}, no.4, 044904 (2017)

\bibitem{STAR:2019bjj}
J.~Adam \textit{et al.} [STAR],
Phys. Rev. C \textbf{102}, no.3, 034909 (2020)


\bibitem{supp:ref}

See Supplemental Material at \href{http://link.aps.org/supplemental/10.1103/PhysRevLett.130.202301}{http://link.aps.org/supplemental/10.1103/PhysRevLett.130.202301}, which include Refs. [\cite{ParticleDataGroup:2018ovx} - \cite{ALICE:2013mez}] for proton feed-down correction results.





\bibitem{STAR:2017ieb}
L.~Adamczyk \textit{et al.} [STAR],
Phys. Rev. Lett. \textbf{121}, no.3, 032301 (2018)

\bibitem{Vovchenko:2019pjl}
V.~Vovchenko and H.~Stoecker,
Comput. Phys. Commun. \textbf{244}, 295-310 (2019)


\bibitem{Shen:2017bsr}
C.~Shen and B.~Schenke,
Phys. Rev. C \textbf{97}, no.2, 024907 (2018)

\bibitem{Denicol:2018wdp}
G.~S.~Denicol, C.~Gale, S.~Jeon, A.~Monnai, B.~Schenke and C.~Shen,
Phys. Rev. C \textbf{98}, no.3, 034916 (2018)

\bibitem{Ropke:2008qk}
G.~Ropke,
Phys. Rev. C \textbf{79}, 014002 (2009)



\bibitem{ParticleDataGroup:2018ovx}
M.~Tanabashi \textit{et al.} [Particle Data Group],
Phys. Rev. D \textbf{98}, no.3, 030001 (2018)




\bibitem{STAR:2019vcp}
J.~Adam \textit{et al.} [STAR],
Phys. Rev. C \textbf{101}, no.2, 024905 (2020)

\bibitem{STAR:2007zea}
B.~I.~Abelev \textit{et al.} [STAR],
Phys. Lett. B \textbf{655}, 104-113 (2007)


\bibitem{STAR:2010yyv}
M.~M.~Aggarwal \textit{et al.} [STAR],
Phys. Rev. C \textbf{83}, 024901 (2011)



\bibitem{ALICE:2015ial}
J.~Adam \textit{et al.} [ALICE],
Eur. Phys. J. C \textbf{75}, no.5, 226 (2015)

\bibitem{ALICE:2019dgz}
S.~Acharya \textit{et al.} [ALICE],
Phys. Lett. B \textbf{794}, 50-63 (2019)


\bibitem{ALICE:2019bnp}
S.~Acharya \textit{et al.} [ALICE],
Phys. Lett. B \textbf{800}, 135043 (2020)

\bibitem{ALICE:2019fee}
S.~Acharya \textit{et al.} [ALICE],
Phys. Rev. C \textbf{101}, no.4, 044906 (2020)

\bibitem{ALICE:2013mez}
B.~Abelev \textit{et al.} [ALICE],
Phys. Rev. C \textbf{88}, 044910 (2013)













\end{thebibliography}


\end{document}